\documentclass[a4paper,aps,twocolumn,nofootinbib]{revtex4}
\RequirePackage[colorlinks,hyperindex]{hyperref}
\RequirePackage[english]{babel}
\RequirePackage[latin1]{inputenc}
\RequirePackage[T1]{fontenc}
\RequirePackage{mathrsfs}
\RequirePackage{amsmath}
\RequirePackage{amssymb}
\RequirePackage{amsbsy}
\RequirePackage{color}
\RequirePackage{bm}
\hypersetup{colorlinks=true,breaklinks=true,urlcolor=blue,linkcolor=red}
\begin{document}
\title{\bf{Dirac Theory in Hydrodynamic Form}}
\author{Luca Fabbri\footnote{luca.fabbri@edu.unige.it}}
\affiliation{DIME, Universit\`{a} di Genova, Via all'Opera Pia 15, 16145 Genova, ITALY\\
INFN, Sezione di Genova, Via Dodecaneso 33, 16146 Genova, ITALY}
\date{\today}
\begin{abstract}
We consider quantum mechanics written in hydrodynamic formulation for the case of relativistic spinor fields to study their velocity: within such a hydrodynamic formulation it is possible to see that the velocity as is usually defined can not actually represent the tangent vector to the trajectories of particles. We propose an alternative definition for this tangent vector and hence for the trajectories of particles, which we believe to be new and in fact the only one that is possible. Finally we discuss how these results are a necessary step to take in order to face further problems, like the definition of trajectories for the multi-particle systems.
\end{abstract}
\maketitle
\section{INTRODUCTION}
Quantum mechanics as founded a century ago is one of the most successful theories ever established, and presumably the most useful of all. Based on the Schr\"{o}dinger equation in the 1920s and finalized with the results of Bell and others from the 1960s up to this day \cite{PBR, S, epr, ab, Bell, B, CHSH, KS}, its applications reach out to chemistry and quantum computing, transistors and lasers, condensed states and medicine.

But while it is obvious that we know quantum mechanics, it is also true that we do not understand it. Whereas we can carry out calculations that give us the answers we want to have, we are not capable of seeing what we are doing. The enterprise with which we seek to understand quantum mechanics, exploring for a way to visualize its content, is the long-lasting search for its interpretation.

Along the decades, several interpretations have been proposed to assign a meaning to the quantum mechanical objects. We do not intend to cover all these interpretations, and interested readers might have a look at \cite{C} and references therein to quench their curiosity. However, it is fundamental to point out that among all candidate models, very few are still compatible with all restrictions imposed by Bell-like types of experiments on non-locality and contextuality \cite{PBR, S, epr, ab, Bell, B, CHSH, KS}. One of the earliest, and certainly among the most famous, is bohmian mechanics.

Originally suggested by de Broglie to solve the paradox of the wave-particle duality by assuming that both waves and particles had a simultaneous existence, it was called the pilot wave model because it stipulated that the wave function, solution of the Schr\"{o}dinger equation, was guiding all particles along their trajectories, given in terms of a guidance equation \cite{LdB1, LdB2}. In time, de Broglie was to be relayed by his students and co-workers, most notably Bohm \cite{DB}, so that the name changed to de Broglie-Bohm model, and then simply to Bohm model.

One of the most attractive features of the bohmian formulation was that it allowed for a much easier way to see what quantum mechanics was all about. Quantum mechanics in bohmian form was not only computing results but also giving a meaning to the elements of the theory.

Such a visualization of quantum mechanics received an additional emphasis with the polar decomposition of complex functions, for which bohmian mechanics could be converted into a form that was analogous to a special type of hydrodynamics for an incompressible fluid \cite{TT}.

However, not all was well. Some difficulties of the formulation were discusses by Takabayasi himself precisely in reference \cite{TT}. One of the most important concerned the definition of velocity for particles. Some aspects of this criticism were soon addressed also by Bohm \cite{B-T}.

Just the same, problems seemed to remain. One of the major issues to face when defining velocities is that there is no satisfactory definition in non-relativistic cases. The bohmian mechanics circumvents the problem by defining the velocity as ratio of momentum over mass. While this definition has many problems that can be forgiven within a non-relativistic environment, the exact proportionality given by $\vec{P}\!=\!m\vec{u}$ holds only in total absence of spin.

In presence of spin, the Schr\"{o}dinger equation is modified into the Pauli equation. One more time the corresponding Bohm model is due to Bohm himself \cite{Bohm-Schiller}.

And again, bohmian mechanics in presence of spin can be written in hydrodynamic formulation \cite{Takabayasi1954}.

More generally, in presence of spin as well as relativistic covariance, the Pauli equation is replaced by the Dirac equation, although the corresponding bohmian mechanics was never really written in detail \cite{Holland}. Problems arise due to the fact that for multi-particle systems, covariance requires each particle to have its own proper time while quantum principles demand all particles to have a single temporal parametrization \cite{md, dgnsz}. For single particles, a hydrodynamic formulation was given, but even for such a formulation the relativistic covariance has been reason of some drawbacks. For example, the earliest-known study of the Dirac equation in hydrodynamic form was made in \cite{Yvon1940}, although surprisingly in a non-manifestly covariant manner. When Takabayasi worked on the hydrodynamic formulation of the Dirac equation he did indeed do it in a manifestly covariant treatment \cite{tr1, tr2, Takabayasi1957}, although again relativistic covariance stood in the way while trying to achieve the polar decomposition of spinors. And when in the same years the polar decomposition of spinors was in fact accomplished \cite{jl1, jl2}, no-one thought to use it to investigate the dynamical character of the Dirac equation.

It is only very recently that the polar form of spinors initially found in \cite{jl1, jl2} has been used to polarly decompose the spinorial covariant derivative in its most general form, in which covariance is respected for whatever coordinate system or local frame, even in curved space-times, in \cite{Fabbri:2020ypd}. The subsequent application to the polar decomposition of the Dirac equation has also been given in the case in which the spinor field interacts with gravity, torsion and electrodynamics \cite{Fabbri:2020ypd}. Finally, the hydrodynamic form of the Dirac equation has been used to conduct a general investigation on bohmian mechanics \cite{Fabbri:2022kfr}.

In the following, we will employ this formulation to examine the aforementioned problem regarding the definition of velocity and the guidance equation for particles.
\section{SPINOR FIELDS}
\subsection{Non-relativistic case}
\label{nonrel}
\subsubsection{Gordon decomposition}
We begin now the introduction of the basic elements of quantum mechanics, that is the theory with which we determine the behaviour of a system described by a wave function $\psi$ in general. As this wave function is complex then its adjoint $\psi^{*}$ will also be necessary.

The dynamics is specified by the Schr\"{o}dinger equation
\begin{eqnarray}
&i\partial_{t}\psi\!+\!\frac{1}{2m}\vec{\nabla}\!\cdot\!\vec{\nabla}\psi\!=\!V\psi\label{S}
\end{eqnarray}
with $V$ a real scalar potential added for generality. Then to get the passage to bohmian mechanics we perform the so-called Gordon decomposition, namely we consider the Schr\"{o}dinger equation (\ref{S}) contracted onto $\psi^{*}$ to obtain a scalar equation, split into real and imaginary parts as
\begin{eqnarray}
&\partial_{t}(\psi^{*}\psi)\!+\!\vec{\nabla}\!\cdot\![\frac{i}{2m}(\vec{\nabla}\psi^{*}\psi
\!-\!\psi^{*}\vec{\nabla}\psi)]\!=\!0\label{cont}\\
\nonumber
&\frac{i}{2}(\psi^{*}\partial_{t}\psi\!-\!\partial_{t}\psi^{*}\psi)+\\
&+\frac{1}{2m}[\frac{1}{2}\vec{\nabla}\!\cdot\!\vec{\nabla}(\psi^{*}\psi)
\!-\!\vec{\nabla}\psi^{*}\!\cdot\!\vec{\nabla}\psi]\!=\!V\psi^{*}\psi\label{energ}
\end{eqnarray}
as a little computation would easily show.

Equation (\ref{cont}) is a continuity equation, the one that will be taken, together with the conservation law of the mass, to establish a link between momentum and velocity, and hence the guidance equation. Indeed, we can say that the passage from quantum mechanics to its bohmian form is mathematically implemented by the conversion from the Schr\"{o}dinger equation to its Gordon decomposition.

\subsubsection{Polar formulation}
Because the wave function is in general a complex function it is always possible to have it written as the product of a module times a unitary phase according to 
\begin{eqnarray}
&\psi\!=\!\phi e^{i\alpha}\label{wfpol}
\end{eqnarray}
with $\phi$ and $\alpha$ being real scalar functions.

The passage to the hydrodynamic formulation can be made by substituting the polar form (\ref{wfpol}) into (\ref{cont}-\ref{energ}) getting
\begin{eqnarray}
&\partial_{t}\phi^{2}
\!+\!\vec{\nabla}\!\cdot\!(\phi^{2}\vec{\nabla}\alpha/m)\!=\!0\label{contpolar}\\
&-\partial_{t}\alpha\!+\!\frac{1}{2m}(\phi^{-1}\vec{\nabla}\!\cdot\!\vec{\nabla}\phi
\!-\!\vec{\nabla}\alpha\!\cdot\!\vec{\nabla}\alpha)\!=\!V\label{energpolar}
\end{eqnarray}
respectively. It is straightforward to check that (\ref{contpolar}-\ref{energpolar}) together imply the Schr\"{o}dinger equation (\ref{S}) in general.

Equation (\ref{contpolar}) is the continuity equation (\ref{cont}) written in a simpler form whereas (\ref{energpolar}) is a Hamilton-Jacobi equation once the phase is recognized as the action functional. In this case in fact we can set $-\partial_{t}\alpha\!=\!H$ and $\vec{\nabla}\alpha\!=\!\vec{P}$ so that (\ref{contpolar}) and (\ref{energpolar}) can be interpreted as the continuity equation giving the conservation of the mass and the expression of the energy of the system, provided that $\phi^{2}\!=\!\rho$ with $\rho$ the density of matter, alongside to $\vec{P}\!=\!m\vec{v}$ with $\vec{v}$ the velocity of the material particle, and where the total potential is given by $V$ plus the term $Q\!=\!-(2m\phi)^{-1}\vec{\nabla}\!\cdot\!\vec{\nabla}\phi$ known as quantum potential. According to this view the quantum mechanical treatment has been transformed into a purely classical treatment but with the potential shifted so to contain also a quantum contribution \cite{DB, TT}. This is the essence of bohmian mechanics in its original formulation.

Nowadays, it is preferred not to mention overmuch the quantum potential, or the Hamilton-Jacobi equation, and to employ the formulation consisting of the definition of density $\phi^{2}\!=\!\rho$ with the guidance equation $\vec{P}\!=\!m\vec{v}$ beside the Schr\"{o}dinger equation (\ref{S}) \cite{Hollandbook, Tumulkabook}. Such an approach is therefore the same as considering the system (\ref{contpolar}-\ref{S}). As (\ref{S}) is equivalent to the system (\ref{contpolar}-\ref{energpolar}), the system (\ref{contpolar}-\ref{S}) is equivalent to the system (\ref{contpolar}-\ref{energpolar}), and therefore this version is equivalent to the original version by Bohm. However, the above argument also shows that the version that we employ today contains a certain redundancy with respect to the version that is due initially to Bohm himself.
\subsection{Relativistic case}
\label{rel}
\subsubsection{Gordon decomposition}
In the previous section we have presented bohmian mechanics insisting on the mathematical formulation and in particular on the way in which it could be derived from quantum mechanics. While the reader familiar with the bohmian mechanics should have recognized the essential elements, it may be that the way they were derived is not the one of standard books on this subject \cite{Hollandbook, Tumulkabook}. This circumstance is wanted because, while not standard, the above presentation is the best suited to be extended up to a relativistic treatment, as we will do next.

The extension will at the same time be an enlargement allowing also the inclusion of spin. Hence Clifford matrices have to be defined. The set of Clifford matrices $\boldsymbol{\gamma}^{a}$ is defined to be such that $\{\boldsymbol{\gamma}_{a},\boldsymbol{\gamma}_{b}\}\!=\!2\mathbb{I}\eta_{ab}$ where $\eta_{ab}$ is the Minkowski matrix. Thus $[\boldsymbol{\gamma}_{a},\boldsymbol{\gamma}_{b}]/4\!=\!\boldsymbol{\sigma}_{ab}$ are generators of the complex Lorentz transformations. Also important is the expression $2i\boldsymbol{\sigma}_{ab}\!=\!\varepsilon_{abcd}\boldsymbol{\pi}\boldsymbol{\sigma}^{cd}$ in terms of the Levi-Civita completely antisymmetric pseudo-tensor defining the $\boldsymbol{\pi}$ matrix (this is usually denoted as a gamma with an index five, but this is misleading in $4$-dimensional space-times, and so we use a notation with no index) and whose existence tells that complex Lorentz transformations are reducible. Then $\boldsymbol{\gamma}_{i}\boldsymbol{\gamma}_{j}\boldsymbol{\gamma}_{k}
\!=\!\boldsymbol{\gamma}_{i}\eta_{jk}-\boldsymbol{\gamma}_{j}\eta_{ik}
\!+\!\boldsymbol{\gamma}_{k}\eta_{ij}
\!+\!i\varepsilon_{ijkq}\boldsymbol{\pi}\boldsymbol{\gamma}^{q}$ holds in general. A complex Lorentz transformation will be denoted as $\boldsymbol{\Lambda}$ so that $\boldsymbol{\Lambda}e^{iq\alpha}\!=\!\boldsymbol{S}$ is the most complete spinorial transformation. Objects that for such a transformation transform as $\psi\!\rightarrow\!\boldsymbol{S}\psi$ and $\overline{\psi}\!\rightarrow\!\overline{\psi}\boldsymbol{S}^{-1}$ are called spinors. We have that $\overline{\psi}\!=\!\psi^{\dagger}\boldsymbol{\gamma}^{0}$ holds and with this pair of adjoint spinors we can build the bi-linear quantities
\begin{eqnarray}
&M^{ab}\!=\!2i\overline{\psi}\boldsymbol{\sigma}^{ab}\psi\\
&S^{a}\!=\!\overline{\psi}\boldsymbol{\gamma}^{a}\boldsymbol{\pi}\psi\\
&U^{a}\!=\!\overline{\psi}\boldsymbol{\gamma}^{a}\psi\\
&\Theta\!=\!i\overline{\psi}\boldsymbol{\pi}\psi\\
&\Phi\!=\!\overline{\psi}\psi
\end{eqnarray}
which are all real tensors and linearly independent. They are however not generally independent, and in fact some constraints hold among them, such as
\begin{eqnarray}
&2\boldsymbol{\sigma}^{\mu\nu}U_{\mu}S_{\nu}\boldsymbol{\pi}\psi\!+\!U^{2}\psi=0\\
&i\Theta S_{\mu}\boldsymbol{\gamma}^{\mu}\psi
\!+\!\Phi S_{\mu}\boldsymbol{\gamma}^{\mu}\boldsymbol{\pi}\psi\!+\!U^{2}\psi=0
\end{eqnarray}
coming alongside to
\begin{eqnarray}
&M_{ab}(\Phi^{2}\!+\!\Theta^{2})\!=\!\Phi U^{j}S^{k}\varepsilon_{jkab}\!+\!\Theta U_{[a}S_{b]}
\end{eqnarray}
as well as
\begin{eqnarray}
&M_{ik}U^{i}\!=\!\Theta S_{k}\\
&M_{ik}S^{i}\!=\!\Theta U_{k}
\end{eqnarray}
and
\begin{eqnarray}
&\frac{1}{2}M_{ab}M^{ab}\!=\!\Phi^{2}\!-\!\Theta^{2}
\label{norm2}\\
&\frac{1}{4}M_{ab}M_{ij}\varepsilon^{abij}\!=\!2\Theta\Phi
\label{orthogonal2}
\end{eqnarray}
and
\begin{eqnarray}
&U_{a}U^{a}\!=\!-S_{a}S^{a}\!=\!\Theta^{2}\!+\!\Phi^{2}\label{norm1}\\
&U_{a}S^{a}\!=\!0\label{orthogonal1}
\end{eqnarray}
as is straightforward to prove and called Fierz identities.

With spin connection $C^{ab}_{\phantom{ab}\mu}$ and gauge potential $A_{\mu}$ it is possible to construct
\begin{eqnarray}
&\boldsymbol{C}_{\mu}
=\frac{1}{2}C^{ab}_{\phantom{ab}\mu}\boldsymbol{\sigma}_{ab}
\!+\!iqA_{\mu}\boldsymbol{\mathbb{I}}\label{spinorialconnection}
\end{eqnarray}
as the spinorial connection. With it
\begin{eqnarray}
&\boldsymbol{\nabla}_{\mu}\psi\!=\!\partial_{\mu}\psi
\!+\!\boldsymbol{C}_{\mu}\psi\label{spincovder}
\end{eqnarray}
is the spinorial covariant derivative of the spinor field.

The dynamics is given by the Dirac equation
\begin{eqnarray}
&i\boldsymbol{\gamma}^{\mu}\boldsymbol{\nabla}_{\mu}\psi
\!-\!XW_{\mu}\boldsymbol{\gamma}^{\mu}\boldsymbol{\pi}\psi\!-\!m\psi\!=\!0
\label{D}
\end{eqnarray}
with $W_{\mu}$ the axial-vector torsion and $X$ its coupling constant, included to be in the most general case where the torsion-spinor interaction is allowed. If we multiply (\ref{D}) by all the linearly independent $\mathbb{I}$, $\boldsymbol{\gamma}^{a}$, $\boldsymbol{\sigma}^{ab}$, $\boldsymbol{\gamma}^{a}\boldsymbol{\pi}$, $\boldsymbol{\pi}$ and by $\overline{\psi}$ splitting real and imaginary parts gives
\begin{eqnarray}
&\nabla_{\mu}U^{\mu}\!=\!0\label{divU}\\
&\frac{i}{2}(\overline{\psi}\boldsymbol{\gamma}^{\mu}\boldsymbol{\pi}\boldsymbol{\nabla}_{\mu}\psi
\!-\!\boldsymbol{\nabla}_{\mu}\overline{\psi}\boldsymbol{\gamma}^{\mu}\boldsymbol{\pi}\psi)
\!-\!XW_{\sigma}U^{\sigma}\!=\!0\label{Lodd}\\
\nonumber
&\nabla^{[\alpha}U^{\nu]}\!+\!i\varepsilon^{\alpha\nu\mu\rho}
(\overline{\psi}\boldsymbol{\gamma}_{\rho}\boldsymbol{\pi}\!\boldsymbol{\nabla}_{\mu}\psi\!-\!\!
\boldsymbol{\nabla}_{\mu}\overline{\psi}\boldsymbol{\gamma}_{\rho}\boldsymbol{\pi}\psi)-\\
&-2XW_{\sigma}U_{\rho}\varepsilon^{\alpha\nu\sigma\rho}\!-\!2mM^{\alpha\nu}\!=\!0\label{curlU}
\end{eqnarray}
\begin{eqnarray}
&\nabla_{\mu}S^{\mu}\!-\!2m\Theta\!=\!0\label{divS}\\
&\frac{i}{2}(\overline{\psi}\boldsymbol{\gamma}^{\mu}\boldsymbol{\nabla}_{\mu}\psi
\!-\!\boldsymbol{\nabla}_{\mu}\overline{\psi}\boldsymbol{\gamma}^{\mu}\psi)
\!-\!XW_{\sigma}S^{\sigma}\!-\!m\Phi\!=\!0\label{Leven}\\
\nonumber
&\nabla^{\mu}S^{\rho}\varepsilon_{\mu\rho\alpha\nu}
\!+\!i(\overline{\psi}\boldsymbol{\gamma}_{[\alpha}\!\boldsymbol{\nabla}_{\nu]}\psi
\!-\!\!\boldsymbol{\nabla}_{[\nu}\overline{\psi}\boldsymbol{\gamma}_{\alpha]}\psi)+\\
&+2XW_{[\alpha}S_{\nu]}\!=\!0\label{curlS}
\end{eqnarray}
\begin{eqnarray}
\nonumber
&i(\overline{\psi}\boldsymbol{\nabla}^{\alpha}\psi
\!-\!\boldsymbol{\nabla}^{\alpha}\overline{\psi}\psi)
\!-\!\nabla_{\mu}M^{\mu\alpha}-\\
&-XW_{\sigma}M_{\mu\nu}\varepsilon^{\mu\nu\sigma\alpha}\!-\!2mU^{\alpha}\!=\!0
\label{vr}\\
\nonumber
&(\boldsymbol{\nabla}_{\alpha}\overline{\psi}\boldsymbol{\pi}\psi
\!-\!\overline{\psi}\boldsymbol{\pi}\boldsymbol{\nabla}_{\alpha}\psi)
\!-\!\frac{1}{2}\nabla^{\mu}M^{\rho\sigma}\varepsilon_{\rho\sigma\mu\alpha}+\\
&+2XW^{\mu}M_{\mu\alpha}\!=\!0\label{ai}
\end{eqnarray}
\begin{eqnarray}
\nonumber
&\nabla_{\alpha}\Phi
\!-\!2(\overline{\psi}\boldsymbol{\sigma}_{\mu\alpha}\!\boldsymbol{\nabla}^{\mu}\psi
\!-\!\boldsymbol{\nabla}^{\mu}\overline{\psi}\boldsymbol{\sigma}_{\mu\alpha}\psi)+\\
&+2X\Theta W_{\alpha}\!=\!0\label{vi}\\
\nonumber
&\nabla_{\nu}\Theta\!-\!
2i(\overline{\psi}\boldsymbol{\sigma}_{\mu\nu}\boldsymbol{\pi}\boldsymbol{\nabla}^{\mu}\psi\!-\!
\boldsymbol{\nabla}^{\mu}\overline{\psi}\boldsymbol{\sigma}_{\mu\nu}\boldsymbol{\pi}\psi)-\\
&-2X\Phi W_{\nu}\!+\!2mS_{\nu}\!=\!0\label{ar}
\end{eqnarray}
as is direct to see and known as Gordon decompositions.

Notice that (\ref{divU}) is precisely the relativistic continuity equation expressing the divergence of the $U^{a}$ vector. Its curl is instead given by (\ref{curlU}) in the same group. In a very analogous way (\ref{divS}) is the partially conserved axial-vector current of the $S^{a}$ axial-vector. Its curl is instead given by (\ref{curlS}) in the second group. In the third group (\ref{vr}) can be recognized as what we name Gordon decomposition in strict sense. The fourth group does not seem to contain anything immediately recognizable although it will be the most important group of the four. This partition in four groups will be very important for the following because
each group is equivalent to any other group as we will be going to demonstrate at the end of the next section.

\subsubsection{Polar formulation}
We give now the following theorem. Any spinor (such that $|i\overline{\psi}\boldsymbol{\pi}\psi|^{2}\!+\!|\overline{\psi}\psi|^{2}\!\neq\!0$ in general) can always be written in chiral representation according to the polar form
\begin{eqnarray}
&\psi\!=\!\phi\ e^{-\frac{i}{2}\beta\boldsymbol{\pi}}
\ \boldsymbol{L}^{-1}\left(\begin{tabular}{c}
$1$\\
$0$\\
$1$\\
$0$
\end{tabular}\right)\label{spinor}
\end{eqnarray}
for some $\boldsymbol{L}$ with the structure of a spinor transformation, with $\phi$ and $\beta$ being a real scalar and a real pseudo-scalar called module and chiral angle. With it we see that
\begin{eqnarray}
&M^{ab}\!=\!2\phi^{2}(\cos{\beta}u_{j}s_{k}\varepsilon^{jkab}\!+\!\sin{\beta}u^{[a}s^{b]})
\end{eqnarray}
showing that this bi-linear quantity can always be written in terms of
\begin{eqnarray}
&S^{a}\!=\!2\phi^{2}s^{a}\label{s}\\
&U^{a}\!=\!2\phi^{2}u^{a}\label{u}
\end{eqnarray}
being the spin axial-vector and the velocity vector and
\begin{eqnarray}
&\Theta\!=\!2\phi^{2}\sin{\beta}\label{sin}\\
&\Phi\!=\!2\phi^{2}\cos{\beta}\label{cos}
\end{eqnarray}
being the only degrees of freedom of the spinor. Then
\begin{eqnarray}
&2\boldsymbol{\sigma}^{\mu\nu}u_{\mu}s_{\nu}\boldsymbol{\pi}\psi\!+\!\psi=0\label{aux1}\\
&is_{\mu}\boldsymbol{\gamma}^{\mu}\psi\sin{\beta}
\!+\!s_{\mu}\boldsymbol{\gamma}^{\mu}\boldsymbol{\pi}\psi\cos{\beta}\!+\!\psi=0\label{aux2}
\end{eqnarray}
with
\begin{eqnarray}
&u_{a}u^{a}\!=\!-s_{a}s^{a}\!=\!1\\
&u_{a}s^{a}\!=\!0
\end{eqnarray}
are the Fierz identities. The advantage of writing spinors in polar form is that the $8$ real components are rearranged into the configuration for which the $2$ real scalar degrees of freedom remain isolated from the $6$ components that are always transferable into the frame and which are then recognized to be the Goldstone fields of the spinor. While here we have presented it in a modern way, these results about the polar form are already found in \cite{jl1, jl2}.

However, here we will go further. To begin, we notice that it is always possible to write in general
\begin{eqnarray}
&\boldsymbol{L}^{-1}\partial_{\mu}\boldsymbol{L}\!=\!iq\partial_{\mu}\xi\mathbb{I}
\!+\!\frac{1}{2}\partial_{\mu}\xi_{ij}\boldsymbol{\sigma}^{ij}\label{spintrans}
\end{eqnarray}
for some $\partial_{\mu}\xi$ and $\partial_{\mu}\xi_{ij}$ which are in fact the Goldstone fields of the spinor mentioned above. We can now define
\begin{eqnarray}
&\partial_{\mu}\xi_{ij}\!-\!C_{ij\mu}\!\equiv\!R_{ij\mu}\label{R}\\
&q(\partial_{\mu}\xi\!-\!A_{\mu})\!\equiv\!P_{\mu}\label{P}
\end{eqnarray}
which are a real tensor and a gauge-covariant vector, and as they contain the same information of spin connection and gauge potential, they are called space-time and gauge tensorial connections. It is then possible to write
\begin{eqnarray}
&\!\!\!\!\!\!\!\!\boldsymbol{\nabla}_{\mu}\psi\!=\!(-\frac{i}{2}\nabla_{\mu}\beta\boldsymbol{\pi}
\!+\!\nabla_{\mu}\ln{\phi}\mathbb{I}
\!-\!iP_{\mu}\mathbb{I}\!-\!\frac{1}{2}R_{ij\mu}\boldsymbol{\sigma}^{ij})\psi
\label{decspinder}
\end{eqnarray}
as the spinorial covariant derivative. From it
\begin{eqnarray}
&\nabla_{\mu}s_{i}\!=\!R_{ji\mu}s^{j}\label{ds}\\
&\nabla_{\mu}u_{i}\!=\!R_{ji\mu}u^{j}\label{du}
\end{eqnarray}
are valid as general identities. This is a crucial point, as it is only after that the tensorial connections (\ref{R}-\ref{P}) are defined that the results of Jakobi and Lochak in \cite{jl1, jl2} on the polar form of spinors can be extended to the polar form of the spinorial covariant derivative as done in \cite{Fabbri:2020ypd}.

Being equipped with the polar form of all the bi-linear quantities and of the spinorial covariant derivative we can next proceed to perform the polar formulation of all the Gordon decompositions that are listed in the four groups above. After introducing
\begin{eqnarray}
&E_{\mu}\!=\!\frac{1}{4}\varepsilon_{\mu\rho\alpha\nu}R^{\rho\alpha\nu}\!-\!XW_{\mu}
\!+\!\nabla_{\mu}\beta/2\!+\!ms_{\mu}\cos{\beta}\label{E}\\
&F_{\mu}\!=\!\frac{1}{2}R_{\mu\rho\alpha}g^{\rho\alpha}\!+\!\nabla_{\mu}\ln{\phi}
\!+\!ms_{\mu}\sin{\beta}\label{F}
\end{eqnarray}
they are given by 
\begin{eqnarray}
&F_{\mu}u^{\mu}\!=\!0\label{A1}\\
&E_{\mu}u^{\mu}\!+\!P_{\mu}s^{\mu}\!=\!0\label{A2}\\
&\varepsilon^{\alpha\nu\mu\rho}E_{\mu}u_{\rho}\!+\!F^{[\alpha}u^{\nu]}
\!+\!\varepsilon^{\alpha\nu\mu\rho}P_{\mu}s_{\rho}\!=\!0\label{A3}
\end{eqnarray}
\begin{eqnarray}
&F_{\mu}s^{\mu}\!=\!0\label{B1}\\
&E_{\mu}s^{\mu}\!+\!P_{\mu}u^{\mu}\!=\!0\label{B2}\\
&\varepsilon^{\alpha\nu\mu\rho}E_{\mu}s_{\rho}\!+\!F^{[\alpha}s^{\nu]}
\!+\!\varepsilon^{\alpha\nu\mu\rho}P_{\mu}u_{\rho}\!=\!0\label{B3}
\end{eqnarray}
\begin{eqnarray}
&F_{\mu}u_{j}s_{k}\varepsilon^{jk\mu\alpha}\!+\!E_{\mu}u^{[\mu}s^{\alpha]}\!-\!P^{\alpha}\!=\!0
\label{forwardmomentum}\\
&F_{\mu}u^{[\mu}s^{\alpha]}\!-\!E_{\mu}u_{j}s_{k}\varepsilon^{jk\mu\alpha}\!=\!0
\label{complmomentum}
\end{eqnarray}
\begin{eqnarray}
&F_{\mu}\!-\!P^{\rho}u^{\nu}s^{\alpha}\varepsilon_{\mu\rho\nu\alpha}\!=\!0\label{auxF}\\
&E_{\mu}\!-\!P^{\iota}u_{[\iota}s_{\mu]}\!=\!0\label{auxE}
\end{eqnarray}
respectively. Notice that each group consists of a total of $8$ independent equations, as many as the $8$ real equations, or $4$ complex equations, constituted by the original Dirac equations. The aforementioned fact that every group of equations (\ref{divU}-\ref{Lodd}-\ref{curlU}), (\ref{divS}-\ref{Leven}-\ref{curlS}), (\ref{vr}-\ref{ai}), (\ref{vi}-\ref{ar}) results to be equivalent comes from the fact that each group of equations (\ref{A1}-\ref{A2}-\ref{A3}), (\ref{B1}-\ref{B2}-\ref{B3}), (\ref{forwardmomentum}-\ref{complmomentum}), (\ref{auxF}-\ref{auxE}) results to be equivalent, and this last fact can be demonstrated in a straightforward way using simple tensor algebra. Also easy to prove is the fact that any of the above groups is equivalent to the original Dirac equations (\ref{D}), as it can be demonstrated by considering for instance the last pair (\ref{auxF}-\ref{auxE}) and see that they imply the Dirac equations (\ref{D}) themselves (to this purpose identities (\ref{aux1}-\ref{aux2}) might turn out to be useful). It is important to notice that in being constituted by exactly $8$ equations any group is the most stringent equivalent to the original Dirac equations. It is also important to highlight that the last pair of equations is important because they specify all derivatives of both degrees of freedom given by module and chiral angle. The reader interested in details can find them in reference \cite{Fabbri:2020ypd}.

The continuity equation for the velocity (\ref{A1}) and the analogous continuity equation for the spin (\ref{B1}) are now in the form of a constraint telling that $F_{\mu}$ does not have a projection along velocity and spin, and equations (\ref{complmomentum}) can be seen as a similar constraint for $F_{\mu}$ and $E_{\mu}$ when projected along the planes orthogonal to the velocity and the spin. Because every other equation contains instances of the momentum $P_{\mu}$ it follows that they can all be recognized as relativistic Hamilton-Jacobi equations. Then equation (\ref{forwardmomentum}) is the explicit expression of the momentum itself. Consequently, the objects $F_{\mu}$ and $E_{\mu}$ play the role of relativistic quantum potentials, as it could have been also expected from the fact that they fully specify all the space-time derivatives of both degrees of freedom \cite{Fabbri:2022kfr}.

While the above groups are strictly equivalent to each other, it is also possible to select groups that are equivalent to each other up to some redundancy. They are the (\ref{divU},\ref{divS},\ref{vr},\ref{ar}) and the (\ref{Lodd},\ref{Leven},\ref{ai},\ref{vi}) for the groups having $2$ redundant field equations, then we have that (\ref{divU},\ref{curlU},\ref{vr}), (\ref{Lodd},\ref{curlU},\ref{ai}), (\ref{Lodd},\ref{curlU},\ref{vi}), (\ref{divU},\ref{curlU},\ref{ar}), (\ref{divS},\ref{curlS},\ref{vr}), (\ref{Leven},\ref{curlS},\ref{ai}), (\ref{Leven},\ref{curlS},\ref{vi}), (\ref{divS},\ref{curlS},\ref{ar}) have $3$ redundant field equations, (\ref{curlU},\ref{curlS}) with $4$ redundant field equations, and (\ref{curlU},\ref{vr},\ref{vi}), (\ref{curlU},\ref{ai},\ref{ar}), (\ref{curlS},\ref{vr},\ref{vi}), (\ref{curlS},\ref{ai},\ref{ar}) with $6$ redundant field equations. Albeit being redundant, these field equations may be collected in groups that display a clear meaning and so better for a possible visualization. As an example, the continuity equation is one with an obvious interpretation and the same may be said for the partially-conserved axial-vector current. Beside velocity and spin divergence also their curl can be pictured easily. Two not-so-similar but still comprehensible field equations are those for the momentum. Consequently we end up with the groups of field equations given by (\ref{divU},\ref{curlU},\ref{vr}) and (\ref{divS},\ref{curlS},\ref{vr}), as well as (\ref{curlU},\ref{curlS}), as the field equations with clear visualization and still equivalent (up to redundancies) to one another and each to the Dirac spinor field equations. That this is possible is fundamental for the representation of spinorial fields as a special type of fluid, as for the hydrodynamic form of the fully-relativistic quantum mechanics.
\section{BOHMIAN MECHANICS}
\subsection{Kinematic Velocity}
In the previous section \ref{nonrel} we have considered quantum mechanics in the way first presented by Bohm in the non-relativistic case and in section \ref{rel} we re-derived all by following the same methods in relativistic case. This finishes the work of Yvon \cite{Yvon1940} and Takabayasi \cite{tr1, tr2, Takabayasi1957} by providing the hydrodynamic re-formulation of the Dirac theory in manifest covariance for curvilinear coordinates and local frames, with electrodynamics, torsion and gravity, that is in the most general situation we can have for systems that are constituted by one single particle.

Now, as mention in \ref{nonrel}, the continuity equations (\ref{cont}) or (\ref{contpolar}) can be compared against the continuity equation
\begin{eqnarray}
&\partial_{t}\rho\!+\!\vec{\nabla}\!\cdot\!(\rho\vec{v})\!=\!0\label{mass}
\end{eqnarray}
to infer that
\begin{eqnarray}
&\psi^{*}\psi\!=\!\phi^{2}\!=\!\rho\\
&\frac{i}{2m}(\vec{\nabla}\psi^{*}\psi\!-\!\psi^{*}\vec{\nabla}\psi)
\!=\!\phi^{2}\vec{\nabla}\alpha/m\!=\!\rho\vec{v}
\end{eqnarray}
so that $\phi^{2}\!=\!\rho$ and $\vec{\nabla}\alpha/m\!=\!\vec{v}$ as definitions of density and velocity. The last in fact is simply the guidance condition of the particles. In fact, once the wave function is known, one can employ the module to get the density distribution of the particles and the phase to compute the velocity as the tangent to the trajectories followed by these particles.

Nevertheless, both the Born rule $\phi^{2}\!=\!\rho$ and the guidance condition $\vec{\nabla}\alpha/m\!=\!\vec{v}$ have the same necessity of (\ref{mass}), and albeit this can be justified as conservation law of the mass in differential form, from a theoretical perspective nothing ensures us that such a continuity condition would be valid. Hence, both Born rule and the guidance equation have to be postulated. This is well known \cite{Drezet:2021dyv, Drezet:2020tbu, Drezet:2022sdu}.

It is reasonable to argue that this problem might arise in a non-relativistic environment, because here the velocity does not have a clear definition. So one can think that things might be easier in a relativistic environment, and in fact there appears to be some improvement when relativistic covariance is implemented because in the ensuing $(3\!+\!1)$-dimensional space-time we seem to be capable of defining the velocity of spinors. This is precisely the velocity vector $u^{a}$ as defined by (\ref{u}) in section \ref{rel} above, and it is evidently linked to the definition of relativistic velocity given by Bohm himself \cite{B-T, Holland}. But then again, to an attentive scrutiny, this definition also fails, because such a definition of velocity is genuinely kinematic.

In fact, the velocity vector $u^{a}$ can be computed directly in polar decomposition, where it acquires the form
\begin{eqnarray}
\nonumber
&u^{a}\!=\!\frac{1}{2\phi^{2}}\overline{\psi}\boldsymbol{\gamma}^{a}\psi=\\
\nonumber
&=\frac{1}{2\phi^{2}}\phi\left(1, 0, 1, 0\right)\boldsymbol{L}e^{-\frac{i}{2}\beta\boldsymbol{\pi}}
\boldsymbol{\gamma}^{a}\phi\ e^{-\frac{i}{2}\beta\boldsymbol{\pi}}
\boldsymbol{L}^{-1}\left(\begin{tabular}{c}
$1$\\
$0$\\
$1$\\
$0$
\end{tabular}\right)=\\
\nonumber
&=\frac{1}{2}\left(1, 0, 1, 0\right)\boldsymbol{L}\boldsymbol{\gamma}^{a}
\boldsymbol{L}^{-1}\left(\begin{tabular}{c}
$1$\\
$0$\\
$1$\\
$0$
\end{tabular}\right)=\\
&=\frac{1}{2}\left(1, 0, 1, 0\right)\boldsymbol{\gamma}^{b}\left(\begin{tabular}{c}
$1$\\
$0$\\
$1$\\
$0$
\end{tabular}\right)(\Lambda^{-1})^{a}_{b}\!=\!(\Lambda^{-1})^{a}_{0}
\end{eqnarray}
in which we have used the fact that $\{\boldsymbol{\pi},\boldsymbol{\gamma}_{a}\}\!=\!0$ and that the complex representation $\boldsymbol{L}$ verifies $\boldsymbol{L}^{-1}\boldsymbol{\gamma}^{a}\boldsymbol{L}\!=\!\boldsymbol{\gamma}^{b}(\Lambda)^{a}_{b}$ in terms of its real representation $(\Lambda)^{a}_{b}$ as a well known property of Lie groups. As both degrees of freedom have disappeared, the velocity can always be written as boost coefficients, and so it is a purely kinematic quantity.

The fact that with relativistic covariance we can define objects that are not definable otherwise is also the reason of the fact that such objects are only kinematic.
\subsection{Dynamic Momentum}
That the velocity itself does not contain dynamical information is the consequence of the fact that it is not in general related to the momentum. In the non-relativistic case such a relation is granted by $\vec{P}\!=\!m\vec{v}$ but as we have seen this link must be postulated. So a natural question now is whether a guidance equation should also be postulated in the relativistic case or if in this situation it can be derived from the dynamics. The answer is that it can.

And in fact, we have already seen how. Just consider equation (\ref{forwardmomentum}) written with (\ref{E}-\ref{F}) explicitly. It reads
\begin{eqnarray}
\nonumber
&P^{\alpha}\!=\!m\cos{\beta}u^{\alpha}+\\
\nonumber
&+(\nabla_{\mu}\beta/2\!-\!XW_{\mu}
\!+\!\frac{1}{4}\varepsilon_{\mu\rho\sigma\nu}R^{\rho\sigma\nu})u^{[\mu}s^{\alpha]}+\\
&+(\nabla_{\mu}\ln{\phi}
\!+\!\frac{1}{2}R_{\mu\rho\sigma}g^{\rho\sigma})u_{j}s_{k}\varepsilon^{jk\mu\alpha}
\label{momentum}
\end{eqnarray}
and in it we can recognize in fact the momentum.

Indeed in the case where there are no geometric forces so that $W_{\mu}\!=\!0$ and $R_{\rho\sigma\nu}\!=\!0$ we get
\begin{eqnarray}
\nonumber
&P^{\alpha}\!=\!m\cos{\beta}u^{\alpha}\!+\!\nabla_{\mu}\beta/2u^{[\mu}s^{\alpha]}+\\
&+\nabla_{\mu}\ln{\phi}u_{j}s_{k}\varepsilon^{jk\mu\alpha}
\end{eqnarray}
and if the internal dynamics of the particle can also be neglected then $\beta\!=\!0$ and so
\begin{eqnarray}
&P^{\alpha}\!=\!mu^{\alpha}\!+\!\nabla_{\mu}\ln{\phi}u_{j}s_{k}\varepsilon^{jk\mu\alpha}
\end{eqnarray}
giving the kinematic momentum up to a spin-dependent quantum correction. In the non-relativistic limit
\begin{eqnarray}
&\vec{P}\!=\!m\vec{v}\!+\!\vec{\nabla}\ln{\phi}\!\times\!\vec{s}
\end{eqnarray}
tying orbital to intrinsic angular momentum.

Therefore in the relativistic case it is possible to obtain from the dynamics that type of guidance equation which in the non-relativistic case can only be postulated.
\section{GUIDANCE EQUATION}
An important thing to notice in the previous analysis of the guidance equation is that while in the non-relativistic case the link between momentum and velocity is simplest, in the relativistic case it is much more involved. However, it is still linear, and so we might think about having the momentum inverted in favour of the velocity vector.

To do this, let us introduce the compact forms
\begin{eqnarray}
&\nabla_{\mu}\beta/2\!-\!XW_{\mu}
\!+\!\frac{1}{4}\varepsilon_{\mu\rho\sigma\nu}R^{\rho\sigma\nu}\!=\!Y_{\mu}\\
&-\nabla_{\mu}\ln{\phi}\!-\!\frac{1}{2}R_{\mu\rho\sigma}g^{\rho\sigma}\!=\!Z_{\mu}
\end{eqnarray}
\begin{eqnarray}
&m\cos{\beta}\!-\!Y_{\mu}s^{\mu}\!=\!X
\end{eqnarray}
in terms of which the momentum is
\begin{eqnarray}
&P^{a}\!=\!(X\eta^{ak}\!+\!Y^{k}s^{a}\!+\!Z_{i}s_{j}\varepsilon^{ijka})u_{k}\label{a}
\end{eqnarray}
which can now be inverted. Introducing 
\begin{eqnarray}
&\zeta_{\mu}\!=\!Z_{\mu}/X
\end{eqnarray}
we can finally write
\begin{eqnarray}
\nonumber
&u^{k}\!=\!(1\!+\!\zeta^{2}\!+\!|\zeta\!\cdot\!s|^{2})^{-1}[\eta^{ka}+\\
\nonumber
&+s^{a}s^{k}(1\!+\!|\zeta\!\cdot\!s|^{2})\!+\!\zeta^{a}\zeta^{k}+\\
&+(s^{a}\zeta^{k}\!+\!\zeta^{a}s^{k})\zeta\!\cdot\!s\!+\!\zeta_{i}s_{j}\varepsilon^{ijka}]P_{a}/X
\label{velocity}
\end{eqnarray}
as the expression of velocity in terms of momentum, spin and degrees of freedom of the spinor, as well as the geometric forces. For details, we refer to reference \cite{Fabbri:2019tad}.

In the same case we have seen above with no geometric forces and chiral angle, the reduced expression is
\begin{eqnarray}
&m\!=\!X
\end{eqnarray}
so that
\begin{eqnarray}
\zeta_{\mu}\!=\!-\frac{1}{m}\frac{1}{\phi}\nabla_{\mu}\phi
\end{eqnarray}
which is the quantum potential in relativistic form. This relativistic expression is first-order derivative for exactly the same reason for which the non-relativistic expression is second-order derivative, and that is because the Dirac equations are first-order derivative while the Schr\"{o}dinger equation is second-order derivative, as well known.

Expression (\ref{velocity}) is the explicit form of the velocity in the most general case. Because it contains the expression of the velocity of bohmian mechanics, we regard it as the most general form of the explicit guidance equation.
\section{CONCLUSION}
In this paper, we have considered the spinor representation in polar form first given in \cite{jl1, jl2} and by means of the tensorial connections (\ref{R}-\ref{P}) we have been able to extend such a polar formulation to the general covariant derivative. This has allowed us to provide a corresponding polar decomposition of the Dirac spinorial field equations analogous to the one summarized in \cite{Takabayasi1957} but in the most complete and systematic presentation. In detail, we have seen how the Dirac equations are equivalent to the systems (\ref{divU},\ref{divS},\ref{curlU},\ref{vr}) or (\ref{divU},\ref{divS},\ref{curlS},\ref{vr}) written in terms of the momentum and the curls and divergences of velocity and spin, that is in terms of quantities that are clearly interpreted by means of hydrodynamic analogs, although with some redundancies. However, we have also seen that no redundancy could be found in the systems (\ref{A1}-\ref{A2}-\ref{A3}) or (\ref{B1}-\ref{B2}-\ref{B3}) as well as (\ref{forwardmomentum}-\ref{complmomentum}) and (\ref{auxF}-\ref{auxE}), which were found to be strictly equivalent to one another and each of them was strictly equivalent to the Dirac equations.

We have then seen that in the relativistic case the definition of velocity that was first given in \cite{B-T} is nevertheless void of dynamic information. A non-trivial dynamic definition of velocity is given through the momentum and accordingly we proved that in a relativistic environment it is possible to tie the momentum to the velocity.

Such a guidance equation was given in its most general form displaying manifest relativistic covariance as in (\ref{momentum}) where the momentum has been expressed in terms of the velocity and the spin. Its inverted expression furnishing the velocity in terms of the momentum, and therefore the true guidance equation, has been given in (\ref{velocity}).

Summing up, we have presented the mathematical formalism of the hydrodynamic analog of the Dirac theory, consisting of the tools needed to treat the bohmian form of quantum mechanics in presence of spin and with full relativistic covariance. It is only after having this formalism that the application to multi-particle systems can be attempted in a somewhat meaningful manner.

\

\textbf{Acknowledgment}. I wish to thank Dr. Marie-H\'{e}l\`{e}ne Genest for discussions and financial support.

\

\textbf{Conflict}. The author declares no conflict of interest.

\

\textbf{Funding}. This research received no funding.

\

\textbf{Availability}. No data is available.

\end{document}